\documentclass[11pt,a4paper]{article}

\usepackage{cite}

\usepackage{amsmath, amsthm, amssymb,slashed}
\usepackage{ifpdf}
\ifpdf
  \usepackage[pdftex]{graphicx}
  \usepackage{epstopdf}
\else
  \usepackage[dvips]{graphicx}
\fi
\usepackage{fancyhdr}
\usepackage{mathbbol}
\usepackage{url}
\usepackage{color}
\usepackage{bbm}
\usepackage{dsfont}
\usepackage{bm}
\usepackage{amsfonts}

\def\be{\begin{equation}}
\def\ee{\end{equation}}
\def\bea{\begin{eqnarray}}
\def\eea{\end{eqnarray}}

\parskip=\baselineskip

\def\1{{\bf 1}}
\def\2{{\bf 2}}
\def\3{{\bf 3}}
\def\4{{\bf 4}}

\font\teneurm=eurm10 \font\seveneurm=eurm7 \font\fiveeurm=eurm5
\newfam\eurmfam
\textfont\eurmfam=\teneurm \scriptfont\eurmfam=\seveneurm
\scriptscriptfont\eurmfam=\fiveeurm

\font\teneusm=eusm10 \font\seveneusm=eusm7 \font\fiveeusm=eusm5
\newfam\eusmfam
\textfont\eusmfam=\teneusm \scriptfont\eusmfam=\seveneusm
\scriptscriptfont\eusmfam=\fiveeusm

\font\tencmmib=cmmib10 \skewchar\tencmmib='177
\font\sevencmmib=cmmib7 \skewchar\sevencmmib='177
\font\fivecmmib=cmmib5 \skewchar\fivecmmib='177
\newfam\cmmibfam
\textfont\cmmibfam=\tencmmib \scriptfont\cmmibfam=\sevencmmib
\scriptscriptfont\cmmibfam=\fivecmmib

\begin{document}
\begin{flushright}

\end{flushright}
\vskip 1.5in
\begin{center}
{\bf\Large{
$Y(4260)\to \gamma + X(3872)$ in the diquarkonium picture}}
\vskip 0.5cm {H.-X. Chen$^\P$, L. Maiani$^*$,  A.D. Polosa$^*$, V. Riquer$^*$}

 \vskip 0.05in

 \textit{$^\P$ School of Physics and Nuclear Energy Engineering, Beihang University, Beijing 100191, China and
INFN Laboratori Nazionali di Frascati, Via E. Fermi 40, I-00044 Frascati, Italy} \\
 \textit{$^*$Dipartimento di Fisica and INFN, Sapienza Universit\`a di Roma,\\ P. Aldo Moro 2, I-00185 Roma, Italy}
 \vskip -.4cm

\end{center}
\vskip 0.5in
\baselineskip 16pt
\begin{abstract}
The observed $Y(4260)\to \gamma + X(3872)$ decay is a natural consequence of the diquark-antidiquark description of $Y$ and $X$ resonances. In this note we attempt an estimate of the transition rate, $\Gamma_{\rm rad}$, by a non-relativistic calculation of the electric dipole term of a diquarkonium bound state. We compute  $\Gamma_{\rm rad}$ for generic composition values of the isospin of $X$ and $Y$. Specializing to $I=0$ for $X(3872)$, we find $\Gamma_{\rm rad}= 496$~keV for $Y(4260)$ with $I=0$ and $\Gamma_{\rm rad}=  179$~keV for $I=1$. Combining with BESIII data, we derive upper bounds to  $B(Y\to J/\Psi+\pi+\pi)$ and to $\Gamma(Y\to \mu^+ \mu^-)$. 
We expect to confront these results with  forthcoming data from electron-positron and hadron colliders.
\newline

\end{abstract}

\subsection*{Introduction}
Exotic, hidden charm, mesons known as $X,Y,Z$ resonances have been interpreted in~\cite{noi1,noi07} as tetraquarks, namely states made by two diquark pairs $[cq][\bar c\bar q^\prime]$ with $q,q^\prime$ light quarks.
Each pair is in color $\bm 3$ or $\bar{\bm 3}$ configuration, spin $s,\bar s=1,0$ and relative orbital momentum $L=0,1$. The scheme has met with some degree of success at explaining the rich phenomenology which has emerged from  electron-positron and proton-(anti)proton collider experiments. More information is expected in the future data from LHCb, BES III and Belle II.

The long-standing  conviction, based on consideration of large-$N$ QCD, that tetraquark states could only materialize in the  form of hadronic resonances too broad to be experimentally resolved, has been recently proven incorrect in~\cite{weinberg}. Tetraquarks in large-$N$ QCD have been further studied in~\cite{brodsky}.
The recent discovery of two pentaquark states of opposite parity~\cite{Aaij:2015tga} has reinforced the case for a new spectroscopic series of hadrons, in which diquarks (antidiquarks)  replace antiquarks (quarks) in the classical scheme~\cite{pentapapers}.

Recently, a new paradigm for the spin-spin interactions in hidden-charm tetraquarks has been proposed, which assumes the dominance of spin-spin couplings inside the diquark or the antidiquark~\cite{noi2}. This simple ansatz  reproduces the mass ordering of the three, well identified, spin $1^+$ states, $X(3872)$, $Z(3900)$ and $Z(4020)$ and the pattern of their observed decays. In addition in Ref.~\cite{noi2} the diquark spin assignments of $L=1$ states is discussed, pointing out that $Y(4260)$ has the same spin distribution as $X(3872)$ namely 
\bea
X&=&|0_{cq},1_{\bar c \bar q};L=0\rangle+|1_{cq},0_{\bar c \bar q};L=0\rangle\notag\\
Y&=&(|0_{c q},1_{\bar c \bar q};L=1\rangle+|1_{ c q},0_{\bar c \bar q};L=1\rangle)_{J=1}\label{spindist}
\eea
States are in the basis $|s,\bar s; L\rangle$ where $s$ ($\bar s$) is the diquark (antidiquark) spin and $L$ the relative orbital angular momentum. 

A similar scheme has been extended to exotic, hidden beauty mesons~\cite{ali1}, and shown to give a consistent picture of the decays of $\Upsilon (10890)$ into $\Upsilon(nS)\pi^+\pi^-$ or $h_b(nP)\pi^+\pi^-$, which occur via the intermediate $Z_b, Z_b^\prime$ states~\cite{ali2}.

The suppression of spin-spin interactions between a quark and an antiquark in different diquarks, underlined in~\cite{noi2}, suggests that the overlap of the two constituents is very small, {\it as if} diquark and antidiquark were well separated entities inside the hadron.
In the present paper we pursue this idea to the extreme consequences by considering the approximation where a diquark and an antidiquark can be described as pointlike. $X,Y,Z$ would be, in this case, bound  particle-antiparticle systems,  that we call diquarkonia for brief. 
We shall see that this extremely simplified picture leads to a reasonable approximation to the mass spectrum of $S$ and $P$ wave tetraquarks. 

The diquarkonium picture has been introduced by A. Ali~{\it et al.}~\cite{alipoint} to study the production and decay of the $Y(10890)$ considered as the $b$-tetraquark analog to the $Y(4008)$.
The annihilation of a diquarkonium with 
$s, {\bar s}=0$ has been treated in~\cite{alipoint}  as the annihilation of a pair of spinless, pointlike particles.
The extension to $Y(4260)\to\mu^+ \mu^-$ 
deserves further consideration, given that the diquark and the antidiquark in the $Y$ have not the same spin and the coupling to the photon is not simply determined by the charges.

We study the diquarkonia mass spectrum in the non-relativistic approximation, using the Cornell potential previously applied to charmonia~\cite{corn1,corn2} and then, equipped with the corresponding wave functions, we compute the predicted rate of the ED1 allowed transition with $\Delta L=1,\Delta s=0$
\be
Y(4260) \to \gamma X(3872)\label{process}
\ee
which  arises naturally from (\ref{spindist}).

Using the masses of the identified $X,Y,Z$ states, we find parameters of the potential rather similar to the Cornell parameters and confirm the identification of the $Z(4430)$ as the first radial excitation of $Z(3900)$.

We compute the rates of the radiative transition for isospin $I=0,1$ of $X(3872)$ and $Y(4260)$.
Assuming $X(3872)$ to be an isospin singlet, we find
\bea
&&\Gamma(Y(4260)\to \gamma X(3872))=\nonumber\\ 
&&= 496~{\rm keV}~ (I:~0\to 0)  \label{pointzero}\\
&&= 179~{\rm keV}~(I:~1\to 0) \label{pointone}
\eea
and compare this result to the available experimental information~\cite{besIII}.

The rate of the radiative decay~(\ref{process}) has been computed in Ref.~\cite{hanha} in the molecular scheme, describing $Y(4260)$ and $X(3872)$ as $DD_1$ and $DD^\star$ bound states respectively. The resulting rate turns out to be considerably smaller than the values indicated in~(\ref{pointzero}) or (\ref{pointone}).

\subsection*{Diquark masses}
For $S$-wave states diquarkonia one writes the rest frame Hamiltonian
\be
M(S{\rm -wave})=2M_{cq}+2\kappa_{cq}({\bm s}_c \cdot {\bm s}_q+{ \bm {\bar s}}_c \cdot {\bm { \bar s}}_q)
\label{ss1}
\ee
where ${\bm s}$ ($ \bm{\bar s}$) denotes the quark (antiquark) spin and $M_{cq}$ is the effective diquark mass.
In the  $|s,\bar s\rangle_J$ basis, $S$-wave tetraquarks with $J^P=1^+$ are described~\cite{noi2} by
\bea
&&  J^{P}=1^{+}\quad C=+\quad\quad \quad\quad\quad X_1= \frac{1}{\sqrt{2}}\left(|1,0\rangle _1+|0,1\rangle _1\right)=X(3872)\label{uno++} \\
&&  J^{P}=1^{+}\quad G=+\quad\quad\quad\quad\quad  \left\{\begin{array}{l}Z=\frac{1}{\sqrt{2}}\left(|1,0\rangle _1-|0,1\rangle _1\right)=Z(3900)\\
Z^{\prime}=|1,1\rangle_1=Z(4020) \end{array}\right.
\label{uno+-}
\eea
$M_{cq}$ can be estimated from the $X(3872)$ and $Z(4020)$ masses, subtracting the spin-spin contributions
\bea
&&M(X)=M(Z)=2M_{cq}-\kappa_{cq}\nonumber \\
&&M(Z^\prime)=2M_{cq}+\kappa_{cq}\nonumber \\
&&M_{cq}=\frac{1}{4}~\left( M[Z(3900)]+M[Z(4020)] \right) \approx 1980~{\rm MeV}
\label{2qm}
\eea
As a first approximation, we shall use $M_{cq}$  as input mass in the Schr{\"o}dinger equation that gives the diquarkonia wave functions and masses.

In the case of charmonium, the input charm quark mass in the Schr{\"o}dinger equation is obtained from the leptonic width $\Gamma(J/\Psi \to e^+ e^-)$, see~\cite{corn1}.
In our case, the leptonic width of  $Y(4260)$ is not available yet and we shall be content to use the value (\ref{2qm}) as input. We have verified that the various quantities are little sensitive (only to few percents) to variations of the input diquark mass around this value.

\subsection*{Bound state masses }
The simplest description of diquarkonia is in term of a non-relativistic potential, $V(r)$. For this first exploration we take the  Cornell potential~\cite{corn1,corn2} with one Chromo-Coulombic  and one confining term
\be
V=-A\,\frac{1}{r}+\nu\, r 
\label{cornpot}
\ee

For charmonia, one finds~\cite{corn2}
\be
A=0.47,~\nu=0.19~{\rm GeV}^{2}~({\rm charmonium~spectrum})
\label{cornpar}
\ee

For diquarkonia, we leave the parameters as free variables to be determined by comparison of diquarkonia eigenvalues $1S$,  $2S$ and $2P$, to the mass differences of the $J=1$ states, $X(3872)$ or $Z(3900)$, $Y(4260)$ and $Z(4430)$, subtracted of spin dependent terms. The subtraction is straightforward for the $S$-wave states, but for $P$-waves it requires the determination of not well known spin-orbit couplings~\cite{noi2}, which introduces a non negligible uncertainty.

Let us assume, as in~\cite{noi1,noi2}, that we can write
\bea
&&M(X)= M_0(1S)+ {\rm spin~ interaction~ terms} \\ \nonumber
&&M(Y)= M_0(2P)+ {\rm spin~ interaction~ terms},\\ \nonumber
&&{\rm etc.}
\eea
where in the r.h.s. we have introduced the eigenvalues of the Schr{\"o}dinger equation, $M_0(1S)$, etc.

 Explicitly, spin interaction terms are obtained from a parametrization of the constituent quark Hamiltonian, which generalizes Eq.~(\ref{ss1}) to include orbital angular momentum excitation~\cite{noi2}~\footnote{Signs are chosen so that, for $B_c$, $a$, $\kappa_{qc}$ positive,  energy increases for increasing $\bm L^2$ and $\bm S^2$. As remarked in~\cite{noi2}, this Hamiltonian is not the most general one as it does not include tensor terms which are known to be important in charmonium. The Hamiltonian describes well the $J=1$ states but it could not be reliable for states with higher $J$.}
\begin{equation}
M=M_{00}+B_c\frac{\bm L^2}{2} -2a~\bm L\cdot \bm S +2\kappa_{qc}~\left[(\bm s_q \cdot \bm s_c)+(\bm s_{\bar q} \cdot \bm s_{\bar c})\right]\label{formula1}
\end{equation}

Obvious manipulations lead to
\begin{eqnarray}
&& M=M_{00}+B_c\frac{L(L+1)}{2} + a~\left[L(L+1)+S(S+1) -2\right] +\notag \\
&&+ \kappa_{qc}~\left[s(s+1)+\bar s (\bar s +1)-3\right]
\label{formula2}
\end{eqnarray}
and we read
\bea
&&M(X(3872))=M_{00} - \kappa_{qc}\nonumber \\
&&M(Y(4260))=M_{00} + B_c+ 2 a - \kappa_{qc}\nonumber \\
&& M(Z(4430))=M_{00}^\prime - \kappa_{qc}
\eea
($M_{00}^\prime$ is the analog of $M_{00}$ for the first radial excitation) so that
\bea
&&M_0(1S)=M(X(3872))+\kappa_{qc}\nonumber \\
&&M_0(2P)=M(Y(4260))-2a+\kappa_{qc}\nonumber \\
&& M_0(2S)=M(Z(4430))+\kappa_{qc}
\eea
and
 \bea
&& M_0(2S)-M_0(1S)=M(Z(4430))-M(Z(3900)) \\
&& M_0(2P)-M_0(1S)=M(Y(4260))-M(X(3872))-2a\label{massdiff}
 \eea
\begin{table}
\begin{center}
\begin{tabular}{| l | c | c | c | c |}
\hline
   Diquarkonium & $X$ & $Z$ & $Z^\prime$ & $Y$   \\
   \hline\hline
 $1 S$& $3871.69\pm0.17$ & $3888.7\pm3.4$ & $4023.9\pm2.4$ & $$ \\
  \hline
$2 S$& $$ & $4485^{+40}_{-25} $ & $$& $$ \\
  \hline
$2 P$ & $$ & $$ & $$ & $4251\pm9$ \\
  \hline
\end{tabular}
  \end{center} \caption{\footnotesize Masses of well identified $X,Y,Z$ states used in the text~\cite{pdg}.}\label{tabmass1}
\end{table}

We use the mass values summarized in Tab.~\ref{tabmass1}~\cite{pdg} and take the value $a=73$~MeV from the fit to the masses of $Y$ states in~\cite{noi2}\footnote{see Eq.~(47) there, for the case in which: $Y_3=|(1_{cq},1_{\bar c\bar q})_{S=0};L=1\rangle =Y(4220)$,  the narrow structure in the $h_c$ channel~\cite{cinfis}($S$ is the total tetraquark spin). Identifying $Y_3=Y(4290)$, the broad structure in the $h_c$ channel~\cite{cinfis}, would lead to a result  consistent with $A=0$.} to which we attribute a theoretical error estimated to be not less than $50\%$. We find
\bea
&& M_{2S}-M_{1S}= 0.60\pm 0.03~{\rm GeV}\nonumber \\
&& M_{2P}-M_{1S}=0.23\pm 0.07~{\rm GeV}
\label{massformula}
 \eea

We solve numerically the Schr{\"o}dinger equation~\cite{code} using the diquark mass in (\ref{2qm}).

\begin{figure}[htb!]
\centering
\includegraphics[scale=0.80]{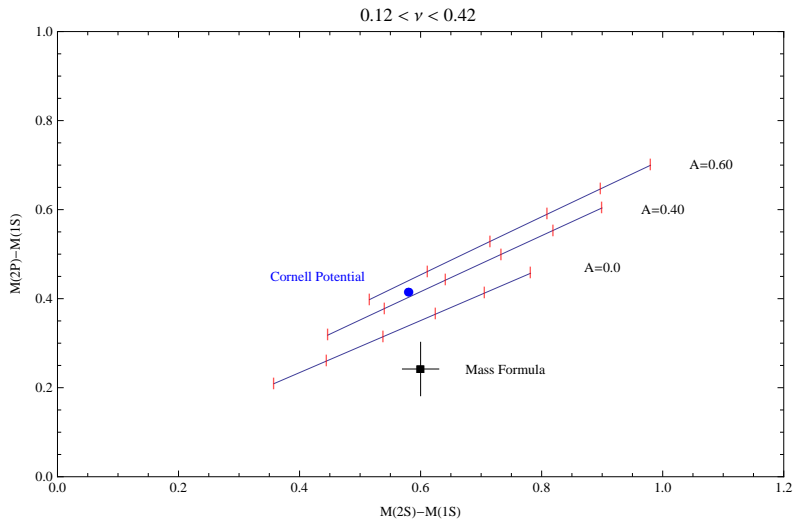}
\caption{\small  Results for the mass-differences in the plane $M(2S)-M(1S)$ and $M(2P)-M(1S)$ (in GeV). The round dot represents the result for the Cornell potential with charmonium parameters given in Eq.~(\ref{cornpar}) and the squared box with errors corresponds to the eigenvalue differences estimated in~(\ref{massformula}). Lines indicate the results computed with fixed $A$ and varying $\nu$. 
}
\label{fig1}
\end{figure}

Results for the mass-differences are reported in Fig.~\ref{fig1}, in the plane of the eigenvalue differences $2S-1S$ and $2P-1S$. The result for the Cornell potential with charmonium parameters is given by the round dot whereas the squared box with errors corresponds to the eigenvalue differences estimated in~(\ref{massformula}). Lines indicate the results computed with fixed $A$ while varying $\nu$. Approximate agreement with the mass formula point is obtained for
 \be
A\sim 0,~\nu=0.25~{\rm GeV}^{2}~({\rm diquarkonium~spectrum})
\label{diqpar}
\ee

The difference $2S-1S$ is  well reproduced for both sets of parameters, Eqs.~(\ref{cornpar}) and~(\ref{diqpar}), reinforcing the case for $Z(4430)$ to be the first radial excitation of $Z(3900)$~\cite{noi07,brodsky}. 
The difference between the parameters in~(\ref{cornpar}) and~(\ref{diqpar}) may be due to the inaccuracy of the mass formula or to the fact that the diquark is not as pointlike as the $c$ quark, therefore less sensitive to the short distance effects embodied by the Coulomb term.

\subsection*{The ED1 transition}
We consider the process
\be
Y(4260)\to \gamma + X(3872)
\label{proc}
\ee
as the ED1 transition from a $P$-wave to a $S$-wave tetraquark with the same spin structure.
Diquarks are taken as pointlike objects of electric charge Q
\be
Q=\left\{\begin{array}{c} +\frac{4}{3}~{\rm for}~[cu] \\ \\ +\frac{1}{3}~{\rm for}~[cd] \end{array}\right.
\ee

The Hamiltonian (radiation gauge) is
\be
H=e Q~ {\bm v}\cdot {\bm A}({\bm x})
\label{hamilt}
\ee
where $\bm A$ is the vector potential, $\bm x$ the coordinate  and $\bm v$ the relative velocity of the particles in the centre of mass system, with the diquark reduced mass
\be
\mu=\frac{1}{2} M_{2q}
\label{redm}
\ee
and $M_{2q}$ given by (\ref{2qm}). In the dipole approximation where we set ${\bm A}({\bm x})\approx {\bm A}({\bm 0})$, the matrix element for the decay is
\bea
{\cal M}_{if}&=& e~\frac{1}{\sqrt{2\omega}}~\langle X,m |Q\,{\bm v}|Y,k\rangle~\cdot {\bm \epsilon}({\bm q})=\\
&=&  ie\omega ~\frac{1}{\sqrt{2\omega}}~\langle X,m |Q\,{\bm x}|Y,k\rangle~\cdot {\bm \epsilon}({\bm q})
\label{matrel}
\eea
where $\bm \epsilon$ and $\bm q$ are the polarization vector and momentum of the photon, $\omega =E_f -E_i$ its energy and $m$ and $k$ label the spin states of $X$ and $Y$ respectively.

The total rate is obtained by~(\ref{matrel})
\bea
\Gamma&=&~e^2 \int \frac{d^3q}{(2\pi)^3 2\omega}\,\omega^2\,(2\pi)\,\delta(E_f-E_i -\omega)~\left(\delta_{ij}-n_i n_j\right) \frac{1}{3}\,\sum_{m,k}\langle Q\, x^i\rangle ~ \langle Q\, x^j\rangle^\star= \nonumber \\
&=&\frac{4\,\alpha\,\omega^3}{9}  \sum_{m,k,i}|\langle Q\,x^i \rangle|^2
\label{rategen}
\eea
where we used 
\be
\int\,d\Omega\,(\delta_{ij}-n_in_j)=\frac{2}{3}(4\pi)\delta_{ij}
\label{solidom}
\ee
with $n^i=q^i/\omega$.

\subsection*{Diquarkonium wave-functions and transition radius}
Consider first diquarkonia with a given flavor composition, {\it e.g.}  $Y_u=[cu][\bar c \bar u]$ or $Y_d=[cd][\bar c \bar d]$.
In the non-relativistic approximation, 
state vectors corresponding to $Y$ ($P$-wave) or  $X$ ($S$-wave) are written as
\bea
&& N_Y \langle Y,k|=\langle 0|\,\int d^3x~R^{2P}(r)\frac{x^i}{r}\epsilon_{ijk} \, \left[d^j_a\left(\frac{{\bm x}}{2}\right)(d_c)^a\left(-\frac{{\bm x}}{2}\right)+d_a\left(\frac{{\bm x}}{2}\right)(d_c)^{aj}\left(-\frac{{\bm x}}{2}\right)\right]\\
\label{Ybra}
&& N_X \langle X,m|=\langle 0|\,\int d^3x~R^{1S}(r)\, \left[d^m_a\left(\frac{{\bm x}}{2}\right)(d_c)^a\left(-\frac{{\bm x}}{2}\right)+d_a\left(\frac{{\bm x}}{2}\right)(d_c)^{a,m}\left(-\frac{{\bm x}}{2}\right)\right]
\label{Xbra}
\eea
where $d$ and $d_c$ (or  $d^m$ and $d_c^m$) are destruction operators of diquark and antidiquark with spin $S=0$ ($S=1$) and $R(r)$ the radial wave functions. We have made explicit the color index $a=1,2,3$. The normalization factors are obtained from (non-relativistic) identities of the form
\be
\langle 0|d_a^j({\bm x})[d_b^l({\bm y})]^\dagger |0\rangle=\delta_a^b~\delta^{jl}~\delta^{(3)}({\bm x}-{\bm y}),~{\rm etc.}
\ee
to wit
\bea
&& N_Y^2= 2^6 (2N)\,\frac{2}{3}\,(4\pi) \,\delta_{k k^\prime}~\delta^{(3)}(0)\\ 
\label{normy}
&& N_X^2=2^6 (2N)\,(4\pi)\,\delta_{m m^\prime}~\delta^{(3)}(0)
\label{normx}
\eea
where~(\ref{solidom}) has been used and the number of colors is $N=3$.

The transition radius is then computed between normalized states to be
\bea
&&\langle X,m|x^i|Y,k\rangle=\frac{1}{\sqrt{6}}~\epsilon_{mik}~\langle r \rangle\\
\label{transrad}
&&\langle r \rangle= \langle r \rangle_{2P\to 1S}= \frac{\int_0^{\infty} r \left[y^{1S}(r)y^{2P}(r)\right]dr}{\sqrt{\int_0^{\infty}dr\, (y^{1S}(r))^2}\sqrt{\int_0^{\infty}dr\, (y^{2P}(r))^2}}
\eea
and we have introduced the reduced radial wave functions of the $1S$ and $2P$ wave-functions $y(r)=r R(r)$ computed numerically~\cite{code}.

Finally, we consider the general isospin structure of $Y(4260)$ and $X(3872)$, defining
\bea
&&X(3872)=\cos{\theta}\,X_u+\sin\theta\,X_d\notag\\
&& Y(4260)=\cos{\phi}\,Y_u+\sin\phi\,Y_d
\eea
and obtain
\bea
&&\langle X,m | Q\,x^i |Y, k\rangle =\frac{1}{\sqrt{6}}~\epsilon_{mik}~Q_{\rm eff}~\langle r \rangle\notag\\
&&Q_{\rm eff}=\left(\frac{4}{3}\cos{\theta}\cos{\phi} + \frac{1}{3}\sin{\theta}\sin{\phi} \right)
\label{radfin}
\eea

 \subsection*{Diquarkonium rate}
 With~(\ref{rategen}) and~(\ref{radfin}), we obtain
\be
\Gamma(Y(4260)\to \gamma + X(3872))= \frac{4\,\alpha\, \omega^3}{9}~Q_{\rm eff}^2 \,\langle r \rangle^2 = 154.2\times Q_{\rm eff}^2\left(\frac{\langle r \rangle}{{\rm GeV}^{-1}}\right)^2~{\rm keV}
\label{ratefin}
\ee
Note that $0 \leq Q_{\rm eff}^2\leq (4/3)^2$, with zero attained when $Y=Y_{u}$ and $X=X_{d}$ or viceversa and the maximum when $Y$ and $X$ have only $u$-flavor.

As indicated by data, we take $X(3872)$ close to a pure $I=0$ state. For the two sets of parameters of the potential, Eqs.~(\ref{cornpar}) and (\ref{diqpar}), we summarize in Tab.~\ref{tabmass} $(i)$ the numerical values of the transition radius and $(ii)$ the rate for $Y(4260)$ with $I=0,1$. 

With the indicated numerical value of the radius, we are at the border of the dipole approximation, since $\omega\langle r\rangle\sim 0.8$, not so much smaller than one. The situation, however,  is not so different from the radiative transition $\chi_{c2}\to J/\Psi\gamma$ which has $\omega\langle r\rangle \sim 0.86$, with estimated $\sim 10\%$ corrections, see~\cite{volo}.

\begin{table}
\begin{center}
\begin{tabular}{| c | c | c | c | c |}
\hline
   ~~ & charm.~potential,~Eq.~(\ref{cornpar}) & diquark.~potential,~Eq.~(\ref{diqpar}) & $Q_{\rm eff}^2$    \\
   \hline\hline
 $\langle r \rangle$, GeV$^{-1}$& $1.84$ & $2.15$ &  \\
  \hline
$\Gamma(I=0 \to I=0)$, keV& $ 361~(3.0\cdot 10^{-3})$ & $496~(4.1\cdot 10^{-3})$ & $25/36$ \\
\hline
$\Gamma(I=1 \to I=0)$, keV& $ 132~(1.1\cdot 10^{-3})$ & $179~(1.5\cdot 10^{-3})$ & $1/4$ \\
  \hline\hline
\end{tabular}
  \end{center} \caption{ Transition radius and corresponding decay widths for $Y\to \gamma X$. In parenthesis the branching ratio, assuming $\Gamma_{Y(4260)}=120$~MeV~\cite{pdg}.}\label{tabmass}
\end{table}

The result found in Ref.~\cite{besIII} can be stated as
\be
\frac{B(Y\to \gamma X)B(X\to J/\Psi ~\pi \pi)}{B(Y\to J/\Psi~ \pi\pi)}=5\cdot 10^{-3}
\ee
which, assuming~\cite{pdg}
\be
B(X\to J/\Psi ~\pi \pi)\gtrsim 2.6 \cdot 10^{-2}
\ee
becomes
\be
\frac{B(Y \to \gamma X)}{B(Y\to J/\Psi ~\pi\pi)}< 0.2
\ee

Using our result, we predict
\be
B(Y\to J/\Psi ~\pi\pi)> \left\{\begin{array}{c} 2.1\cdot 10^{-2}\quad (I: 0 \to 0) \\ 0.78\cdot 10^{-2}\quad (I: 1 \to 0)\end{array}\right.
\ee

From the value of $\Gamma(Y\to X\gamma)$ we can also estimate $\Gamma(Y\to e^-e^+)$. We use the well known formula for the peak cross section
\be
\sigma(e^-e^+\to X\,\gamma)_{Y}=\frac{12\pi}{m_Y^2}\frac{\Gamma(Y\to X\gamma)\Gamma(Y\to e^-e^+)}{\Gamma(Y\to \mathrm{All})^2}
\ee
with the experimental determination~\cite{bes3y}
\be
\sigma(e^-e^+\to Y(4260) \to X\,\gamma)=\frac{0.33~\mathrm{pb}}{\mathcal{B}(X\to \pi^+\pi^- J/\Psi)}
\ee
and  
the input values in Table~\ref{tabmass1} and~\ref{tabmass}  (diquarkonium potential)
\be
\Gamma(Y\to e^-e^+)\lesssim \frac{226}{(\Gamma(Y\to X\gamma)/\mathrm{keV})}~\mathrm{keV}= \left\{\begin{array}{c} 0.45 \\ 1.26 \end{array}\right.~{\rm keV}
\ee
and
\be
\sigma(e^+e^-\to \mu^+\mu^-)_Y \lesssim \frac{2871}{(\Gamma(Y\to X\gamma)/\mathrm{keV})^2}~\mathrm{pb}= \left\{\begin{array}{c} 0.01 \\ 0.09 \end{array}\right.~{\rm pb}
\ee
for $Y$ isospin equal to $0$ or $1$, respectively. 
 \subsection*{Conclusions}
 We estimated the transition rates $\Gamma(Y(4260)\to \gamma+X(3872))$ under 
 the assumption that $Y$ is a diquarkonium confined by a Cornell-like potential, either isospin singlet or triplet. 
Our results reinforce the case for $Z(4430)$ to be the first radial excitation of $Z(3900)$~\cite{noi07,brodsky}.
Mass differences between states with different orbital excitations computed with a linearly rising potential  and no Chromo-Coulombic term approximately agree with the mass formula derived from the constituent quark model in~\cite{noi2}.
The results obtained, together with upper bound estimates of $B(Y\to J/\Psi ~\pi\pi)$ and of the $Y$ electronic width, can be confronted with  future data, from electron-positron and hadron colliders. 

\subsection*{Acknowledgements}
We thank Qiang Zhao, Xiao-Yan Shen, Chang-Zheng Yuan, Rinaldo Baldini, Simone Pacetti and Monica Bertani  for interesting discussions.
Part of this work was done at IHEP-Beijing and at the Frascati Laboratories of INFN. L.M. and V.R. thank Prof. Yifang Wang and H.X.C. thanks Prof. Pierluigi Campana for hospitality.

\bibliographystyle{unsrt}

\end{document}